%% file: article.tex
\documentclass{easychair}

\usepackage{amssymb}

\newcommand{\tmrsup}[1]{$\mbox{}^\ensuremath{\text{#1}}$}
\newcommand{\kid}[1]{\textbf{\textit{#1}}}
\newcommand{\kwd}[1]{\textbf{\textup{#1}}}
\newcommand{\spc}{\text{ }}
\newcommand{\assign}{:=}

\begin{document}

\title{Deriving a Hoare-Floyd logic for non-local jumps\\
from a formul\ae-as-types notion of control}

\author{Tristan Crolard\\
Emmanuel Polonowski\\
LACL -- Universit\'e Paris-Est, France\\
\url{{crolard,polonowski}@u-pec.fr}\\
}

\authorrunning{T.~Crolard and E.~Polonowski}
\titlerunning{Deriving a Hoare-Floyd logic for non-local jumps}

\maketitle

\begin{abstract}
We derive a Hoare-Floyd logic for non-local jumps and mutable higher-order procedural variables
from a formul\ae-as-types notion of control for classical logic. The main contribution of this work is the design of an imperative dependent type system for non-local jumps which corresponds to classical logic but where the famous {\it consequence rule} is still derivable.
\end{abstract}


\input{paper}


\end{document}

%% file: paper.tex
Hoare-Floyd logics for non-local jumps are notoriously difficult to obtain,
especially in the presence of local mutable variables {\cite{Tennent91}}. As
far as we know, the question of proving the correctness of imperative programs
which combine local mutable {\textit{higher-order procedural variables}} and
non-local jumps has not even been addressed. On the other hand, we know since
Griffin's pioneering work {\cite{Griffin90}} how to prove the correctness of
(higher-order) functional programs with control in direct style, thanks to the
formul{\ae}-as-types interpretation of classical logic.

In {\cite{Crolard10a}}, Chapter 3, we have thus extended the
formul{\ae}-as-types notion of control to imperative programs with
higher-order procedural mutable variables and non-local jumps. Our technique,
which was inspired by Landin's seminal paper {\cite{Landin65}}, consists in
defining an imperative dependent type system {\kwd{ID}} by translation into a
functional dependent type system (which is actually Leivant's {\kwd{ML1P}}
{\cite{Leivant90}}). This imperative language, called
{\textsc{Loop}}\tmrsup{$\omega$}, was defined by the authors in
{\cite{Crolard09}}.

Similarly to {\kwd{ML1P}}, the imperative type system is parametrized by a
first-order signature and an equational system $\mathcal{E}$ which defines a
set of functions in the style of Herbrand-G\"odel. The syntax of imperative
types of {\kwd{ID}} (with dependent procedure types and dependent records) is
the following:
\[ \sigma, \tau \spc \text{: : =} \spc \kwd{\textbf{nat}} (n) \spc | \spc
   \kwd{\textbf{proc}} \spc \forall \vec{\imath} ( \kwd{\textbf{in}} \spc
   \vec{\tau} ; \kwd{\textbf{out}} \spc \vec{\sigma}) \spc | \spc \exists
   \vec{\imath} (\sigma_1, \ldots, \sigma_n) \spc | \spc n = m \]
Typing judgements of {\kwd{ID}} have the form $\Gamma ; \Omega \vdash e :
\psi$ if $e$ is an expression and $\Gamma ; \Omega \vdash s \vartriangleright
\Omega'$ if $s$ is a sequence, where environments $\Gamma$ and $\Omega$
corresponds respectively to immutable and mutable variables. Note that our
type system is \textit{pseudo-dynamic} in the sense that the type of mutable
variables can change in a sequence and the new types are given by $\Omega'$
(as in {\cite{Xi00}}). For instance, here is the typing rule of the
{\kwd{for}} loop:
\[ \dfrac{\Gamma ; \Omega, \vec{x} : \vec{\sigma} [ \kwd{0} / i] \vdash e :
   \kwd{\textbf{nat}} (n) \hspace{2em} \Gamma, y : \kwd{\textbf{nat}} (i) ;
   \vec{x} : \vec{\sigma} \vdash s \vartriangleright \vec{x} : \vec{\sigma} [
   \kwd{s} (i) / i]}{\Gamma ; \Omega, \vec{x} : \vec{\sigma} [ \kwd{0} / i]
   \vdash \kwd{\textbf{for}} \spc y \assign 0 \spc \kwd{\textbf{until}} \spc e
   \spc \{s\}_{\vec{x}} \vartriangleright \vec{x} : \vec{\sigma} [n / i]} \]

\subsection*{Embedding a Hoare-Floyd logic}

It is almost straightforward to embed a Hoare-Floyd logic into {\kwd{ID}}.
Indeed, let us take a global mutable variable, dubbed {\kid{assert}}, and let
us assume that this global variable is simulated in the usual
\textit{state-passing style} (the variable is passed as an explicit
{\kwd{in}} and {\kwd{out}} parameter to each procedure call). Consequently,
any sequence shall be typed with a sequent of the form $\Gamma ; \Omega,
\kid{\textbf{assert}} : \varphi \vdash s \vartriangleright \Omega',
\kid{\textbf{assert}} : \psi$. If we now introduce the usual Hoare notation for
triples (which hides the name of global variable {\kid{assert}}), we obtain
judgments of the form $\Gamma ; \Omega \vdash \{\varphi\} s \vartriangleright
\Omega' \{\psi\}$. Rules very similar to Hoare rules are then derivable: for
instance, the type of {\kid{assert}} corresponds to the invariant in a loop,
and to the type of {\textit{pre}} and {\textit{post}} conditions in a procedure
type. The only rule which is not directly derivable is the well-known
\textit{consequence rule}:
\[ \dfrac{\Gamma, \Omega \vdash \varphi' \Rightarrow \varphi \hspace{2em}
   \Gamma ; \Omega \vdash \{\varphi\} s \vartriangleright \Omega' \{\psi\}
   \hspace{2em} \Gamma, \Omega \vdash \psi \Rightarrow \psi'}{\Gamma ; \Omega
   \vdash \{\varphi' \} s \vartriangleright \Omega' \{\psi' \}} \]
This rule deserves a specific treatment since no proof-term is required for
the proof obligations. However, it is well-known that in intuitionistic logic
the proof of some formulas have no computational content (they are called
\textit{data-mute} in {\cite{Leivant90}}). The consequence rule is thus
derivable if we restrict (without loss of generality) the set of assertions to
data-mute formulas.

\subsection*{Non-local jumps}

The imperative language was then extended in {\cite{Crolard10a}} with labels
and non-local jumps. At the (dependent) type level, this extension (called
{\kwd{ID}}\tmrsup{$c$}) corresponds to an extension from intuitionistic logic
to classical logic. For instance, the following typing rules for labels and
jumps are derivable (where first-class labels are typed by the negation):
\[ \begin{array}{c}
     \dfrac{\Gamma, k : \neg \vec{\sigma} ; \vec{z} : \vec{\tau} \vdash s
     \vartriangleright \vec{z} : \vec{\sigma} \hspace{2em} \Gamma ; \Omega,
     \vec{z} : \vec{\sigma} \vdash s' \vartriangleright \Omega'}{\Gamma ;
     \Omega, \vec{z} : \vec{\tau} \vdash k : \{s\}_{\vec{z}} ; \spc s'
     \vartriangleright \Omega'} \hspace{2em} \dfrac{\Gamma ; \Omega, \vec{z} :
     \vec{\tau} \vdash k : \neg \vec{\sigma} \hspace{2em} \Gamma ; \Omega,
     \vec{z} : \vec{\tau} \vdash \vec{e} : \vec{\sigma}}{\Gamma ; \Omega,
     \vec{z} : \vec{\tau} \vdash \kwd{\textbf{jump}} (k, \vec{e})_{\vec{z}}
     \vartriangleright \vec{z} : \vec{\tau}'}
   \end{array} \]
However, deriving a Hoare-Floyd logic for non-local jumps is not
straightforward since there is no obvious notion of \textit{data-mute}
formula in classical logic (as noted also in {\cite{Makarov06}}), and thus the
consequence rule is in general not derivable. The problem comes from the fact
that, in presence of control operators, the proof-terms corresponding to
proof-obligations may interact with the program. We shall exhibit an example
of such program and we shall present a general solution to this problem which
relies on the distinction between purely functional terms and imperative
procedures (possibly containing non-local jumps).